\documentclass[twocolumn,pra,aps,superscriptaddress,showpacs]{revtex4}
\usepackage{amsmath,amsfonts,amssymb}
\usepackage{graphicx,subfigure}
\usepackage[colorlinks,urlcolor=blue,linkcolor=blue,anchorcolor=blue,citecolor=blue]{hyperref}
\newcommand{\ket}[1]{\left\vert #1 \right\rangle}

\newcommand{\bla}[1]{\left( #1 \right)}

\usepackage{color}

\begin{document}
	
	\title{Entangling distant solid-state spins via thermal phonons}
	
	\author{Puhao Cao}
	
	\author{Ralf Betzholz}
	\email{ralf\_betzholz@hust.edu.cn}
	
	\author{Shaoliang Zhang}
	
	\author{Jianming Cai}
	\email{jianmingcai@hust.edu.cn}
	\affiliation{School of Physics and International Joint Laboratory on Quantum Sensing and Quantum Metrology, Huazhong University of Science and Technology, Wuhan, China, 430074}	
	

	\begin{abstract}
		The implementation of quantum entangling gates between qubits is essential to achieve scalable quantum computation. Here, we propose a robust scheme to realize an entangling gate for distant solid-state spins via a mechanical oscillator in its thermal equilibrium state. By appropriate Hamiltonian engineering and usage of a protected subspace, we show that the proposed scheme is able to significantly reduce the thermal effect of the mechanical oscillator on the spins. In particular, we demonstrate that a high entangling gate fidelity can be achieved even for a relatively high thermal occupation. Our scheme can thus relax the requirement for ground-state cooling of the mechanical oscillator, and may find applications in scalable quantum information processing in hybrid solid-state architectures.		
	\end{abstract}
	\date{\today}	
	\maketitle
	
	\section{Introduction}
	Quantum entanglement is one of  the most fascinating aspects of quantum theory~\cite{Einstein35,Schrödinger35,Ryszard09}. Moreover, it is also a valuable resource for various types of quantum technologies, such as quantum communication~\cite{Artur91,Naik00,Tittel00,Jonathan05,Ursin07}, quantum metrology~\cite{Giovannetti06} and quantum computation~\cite{Nielsen02,Duan01,Michael02}. The generation of quantum entanglement is thus an essential task in quantum information processing. In particular, it is exceedingly challenging to entangle qubits at a large distance, e.g. due to the lack of strong interactions. For example, the direct coupling between two solid-state spins at a distance on the order of micrometers via their magnetic dipole-dipole interaction is negligible given the constraint of their coherence time. The ability of entangling two such distant qubits thereby becomes an important goal to pursue on the road towards the implementation of large scale quantum information processing.
	
	Hybrid optomechanical and atomic systems offer an appealing way to entangle distant atomic qubits in solid-state systems~\cite{Rogers14}. The essential idea is to realize indirect coupling between quantum systems via the vibrational mode of a mechanical oscillator, which serves as a mediator. For example, the interaction between nitrogen-vacancy (NV) center spin qubits in diamond and a cantilever can be induced by strain or a magnetic field gradient~\cite{Rabl09,Rabl10,Arcizet11,Shimon11,Yin15}. The idea can also be generalized to novel types of mechanical oscillators, such as graphene sheets and nanotubes~\cite{Peng16}. In previous proposals, it is usually assumed that the mechanical oscillator is cooled down close to its motional ground state~\cite{Rogers14,Rabl09,Rabl10,Arcizet11,Shimon11,Yin15,Peng16}. Although it is in general possible to reach average phonon numbers as low as $\bar{n}\approx 0.2$, e.g., by resolved side-band cooling~\cite{Wilson07,Jasper11,Schliesser06,Schliesser08,Teufel11}, it imposes stringent requirements and a large experimental overhead to achieve such a cooling efficiency. Unless the mechanical oscillator is cooled to its ground state the thermal effects of its motion will inevitably degrade the fidelity of the entangling gate between distant two-level quantum systems and severely limit its scalability. We remark that the celebrated S{\o}rensen-M{\o}lmer gate provides a way to achieve robust coupling between trapped ions in thermal motion based on a laser configuration with specific tunings~\cite{Sor99}.
	
	To overcome this challenge, here we propose to engineer a decoherence-free subspace (DFS) with suitable driving fields~\cite{Cai15} to realize an entangling gate between distant solid-state spins. The logical qubit, which is encoded in the DFS of two quantum systems, becomes much less susceptible to the phonon number of the mechanical oscillator. We theoretically demonstrate that the thermal effect from the mechanical oscillator can be significantly suppressed, and that a high fidelity entangling gate can be realized even for a relatively high average phonon number of the mechanical oscillator.  The present result applies to general hybrid systems composed of atomic systems and mechanical oscillators, including nitrogen-vacancy centers in diamond and cantilevers (or nanotubes). Our proposal may thereby find applications in both the generation of large scale entangled states for quantum metrology and scalable quantum computation.
	
	\section{Model}
	The system we consider consists of solid-state spins, which are in close proximity to a mechanical oscillator. The oscillator's frequency is given by $\nu$ and the displacement from its equilibrium position is denoted by $z$. The two-level spins, labeled by $k$, have ground states $|g\rangle_k$, excited states $|e\rangle_k$, and an additional auxiliary state $|a\rangle_k$. Furthermore, an external field, detuned from individual spin resonances by $\Delta_k$, drives the spin transitions with a Rabi frequency $\Omega_k$. A schematic picture of the system is shown in Fig.~\ref{Fig.lab.1}.  
	\begin{figure}[t!]
		\begin{center}
			(a)\hspace{-2mm}\vtop{\vskip+1ex\hbox{\includegraphics[width=0.65\linewidth]{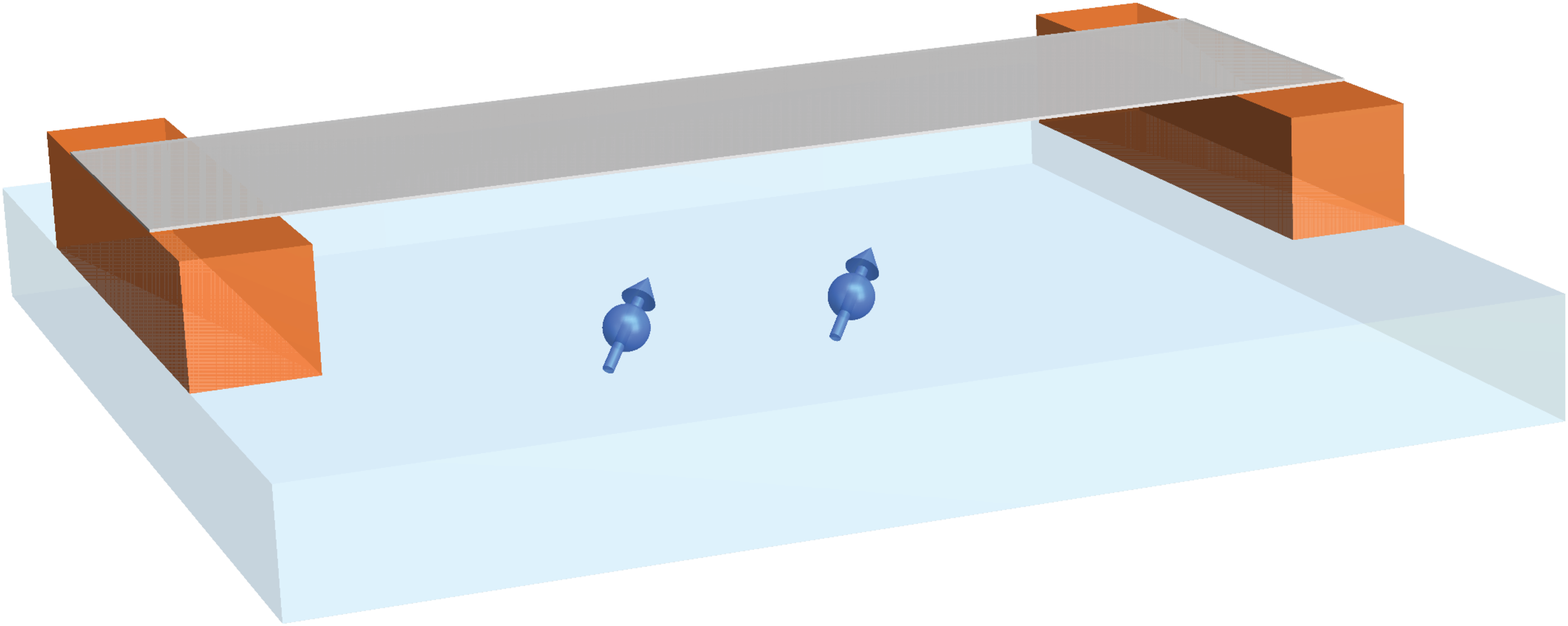}}}\hspace{2mm}
			(b)\hspace{-2mm}\vtop{\vskip-1ex\hbox{\includegraphics[width=0.25\linewidth]{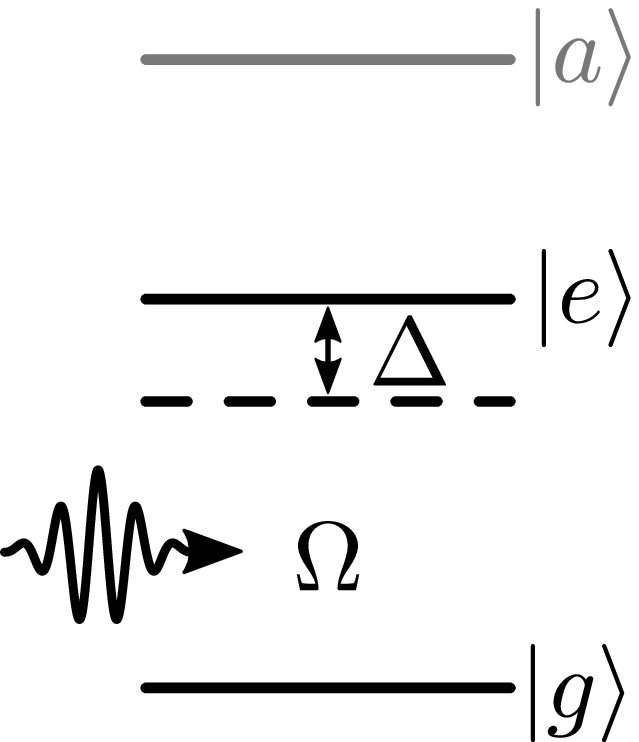}}}
			\label{Fig.lab.1}
		\end{center}
		\caption{(a) Coupling of two distant quantum systems (e.g., NV centers in diamond) via a mechanical oscillator (e.g., a graphene sheet). (b) The energy levels of the spin systems driven by a field $\Omega$ with detuning $\Delta$.}
		\label{Fig.lab.1}
	\end{figure}
	
	The motion of the oscillator induces a position-dependent magnetic field acting on the spin system, thereby coupling the vibrational and the spin degrees of freedom. The total Hamiltonian of the hybrid system, in the frame rotating with the driving field frequency, can be written as~\cite{Rabl09}
	\begin{equation}
	H = H_{\rm A} + H_{\rm M} + H_{\rm I}.
	\end{equation}
	Here, the atomic spin Hamiltonian $H_{\rm A}$, here written without the auxiliary state, is given by 
	\begin{gather}
	H_{\rm A}=\sum_k\frac{\hbar\Delta_k}{2}\sigma_z^{(k)}+\sum_k\frac{\hbar\Omega_k}{2}\sigma_x^{(k)}
	\end{gather}
	with the Pauli operators $\sigma_z^{(k)}=|e\rangle_k\langle e|-|g\rangle_k\langle g|$ and $\sigma_x^{(k)}=|e\rangle_k\langle g|+|g\rangle_k\langle e|$. The mechanical oscillator Hamiltonian $H_{\rm M}$ reads
	\begin{gather}
	H_{\rm M}=\hbar\nu a^\dagger a
	\end{gather}
	with the annihilation and creation operators $a$ and $a^\dagger$, respectively. In a linear approximation the interaction $H_{\rm I}$ between the magnetic field emanating from the mechanical oscillator and the spin system is of the form 
	\begin{gather}
	H_{\rm I}=\sum_k\hbar g_k (a+a^\dagger) \sigma_z^{(k)},
	\end{gather}
	where the coupling constants $g_k$ are proportional to the field gradient along the $z$ direction evaluated at the distance between the oscillators equilibrium position and the $k$th spin. In the next step, we exchange the spin basis by the transform $\sigma_z\leftrightarrow\sigma_x$ and perform a rotating wave approximation to obtain a Tavis-Cummings type Hamiltonian~\cite{Tavis68} of the form
	\begin{align}
	\label{eq:Tavis-Cummings}
	H =&\sum_k\frac{\hbar\Omega_k}{2}\sigma_z^{(k)}+\sum_k\frac{\hbar\Delta_k}{2}\sigma_x^{(k)}+ \hbar\nu a^\dagger a\nonumber\\
	&+ \sum_k\hbar g_k\left[a\sigma_+^{(k)} + a^\dagger\sigma_-^{(k)}\right],
	\end{align}
	where we introduced the atomic raising and lowering operators $\sigma_\pm^{(k)}=|\pm\rangle_k\langle\mp|$, respectively. Here, the states $|\pm\rangle=[|e\rangle\pm|g\rangle]/\sqrt{2}$ denote the eigenstates of the new $\sigma_{z}$, i.e. they are in fact the eigenstates of $\sigma_{x}$ in the original picture. 
	
	Before we continue to elaborate the details of our scheme, we briefly comment on a possible experimental realization of this system. As solid-state spins one 
	could use NV centers in diamond. In this case, one can realize the 
	required level scheme in the ground-state triplet with the states 
	$|g\rangle=|m_s=-1\rangle$, $|e\rangle=|m_s=+1\rangle$, and 
	$|a\rangle=|m_s=0\rangle$. Here, a static magnetic field along the 
	NV-center's $z$ axis can be used to tune the energy of the states,
	the $|g\rangle$-$|e\rangle$ transition could be driven in a 
	Raman configuration, and optical spin polarization allows for an easy preparation of the state $|a\rangle$. The mechanical oscillator on the other hand can be realized, for example, by a graphene sheet carrying a direct current, thereby providing the magnetic field, which couples to the spins, see Fig.~\ref{Fig.lab.1}(a). We note that in our model we assume that the spatial distribution of spins and the mode shapes of the vibrations are such that the spins only couple strongly to a single mode among the manifold of vibrational modes present in such a geometry~\cite{Arash11,Arghavan12}, allowing for a single-mode approach. This kind of architecture would thereby allow for coupling in a two-dimensional array, contrary to realizations with nanotubes. For exemplary experimental parameters we refer to the discussion in Sec.~\ref{sec:conclusion}.
	
	\section{Effective spin-spin coupling}
	The system introduced above allows to achieve a coupling between the otherwise uncoupled spins which is mediated by the mechanical oscillator~\cite{Sor99,Bennett13}. This coupling is best accomplished in the dispersive regime~\cite{Blais04,Govia16}, i.e., the far off-resonant regime where the detuning of the energy splitting of the spins from the oscillator frequency is much larger than their coupling to the oscillator. Considering the Hamiltonian~\eqref{eq:Tavis-Cummings} this is the case for $g_k/\delta_k\ll 1$, with the pseudo-detuning $\delta_k=\Omega_k-\nu$.
	
	\subsection{Coupling two spins}
	In this regime one can derive an effective Hamiltonian in different orders of $g_k/\delta_k$~\cite{James07,Zueco09,Bennett13}. We restrict ourselves to the case of two spins, and the same perturbative treatment in the small parameter $\Delta_k/\Omega_k$ can be applied. In addition to the zeroth-order Hamiltonian $H_0=\sum_k\hbar\delta_k\sigma_z^{(k)}/2$, we thereby obtain the second-order effective Hamiltonian
	\begin{align}
	\label{eq:effective}
	H_{\rm eff}^{(2)}=&\sum_{k=1,2}\left[\frac{\hbar\bar\Delta_k}{2} + \frac{\hbar g_k^2}{2\delta_k}(2a^\dagger a+1)\right]\sigma_{z}^{(k)}\nonumber\\
	&+\frac{\hbar g_1g_2(\delta_1+\delta_2)}{2\delta_1 \delta_2}\left[\sigma_+^{(1)}\sigma_-^{(2)}+
	\sigma_-^{(1)}\sigma_+^{(2)}\right].
	\end{align}
	More details on the derivation can be found in Appendix~\ref{app:effective}. Here, the first term with $\bar{\Delta}_k=\Delta_k^2/2\Omega_k$ is the usual ac Stark shift due to the driving fields, where the role of detuning and Rabi frequency have been interchanged because of the change of basis from $\sigma_z$ to $\sigma_x$. The second term constitutes a phonon-number dependent Stark shift originating in the spin-oscillator interaction. As can be seen from this second term in the Hamiltonian, fluctuations of the phonon number induce an energy fluctuation of the spins and thereby cause dephasing.
	
	Taking into account the non-uniform coupling $g_k$, one can always choose suitable values of $\delta_k$ such that $g_1^2/\delta_1=g_2^2/\delta_2$ is fulfilled by appropriately adjusting the Rabi frequencies. In the same manner, one could also ensure $\bar\Delta_1=\bar\Delta_2$ by tuning the resonance frequencies of the spins or the driving field frequency. Without loss of generality, we therefore assume $g_k=g$ and $\delta_k=\delta$. Assuming for the moment a resonant drive of the spins, viz. $\bar\Delta_k=0$, this leads to the effective Hamiltonian
	\begin{align}
	\label{eq:second_order}
	H_{\rm eff}^{(2)}
	=\alpha^2\hbar\delta\left[\frac{1}{2}(2a^\dagger a+1)S_{z}+\sigma_+^{(1)}\sigma_-^{(2)}+
	{\sigma_-^{(1)}\sigma_+^{(2)}}\right]
	\end{align}
	with the operator $S_z=\sigma_{z}^{(1)}+\sigma_{z}^{(2)}$ and the smallness parameter $\alpha=g/\delta$.
	
	In order to overcome the phonon-induced dephasing mentioned above, we consider the subspace spanned by the two states $|0\rangle=\ket{-}_1\ket{+}_2$ and $|1\rangle=\ket{+}_1\ket{-}_2$. The fact that the operator $S_z$ annihilates both these states, viz. $S_z|0\rangle=S_z|1\rangle=0$, warrants that up to second order in $\alpha$ this subspace is fully protected from the influence of phonon number fluctuations, i.e., up to this order the dynamics in this subspace is independent of the temperature of the oscillator and one finds effective oscillations between $|0\rangle$ and $|1\rangle$ which are mediated by the oscillator.
	
	The residual effects of the phonon number arise only from higher order terms in the effective Hamiltonian.  The explicit form of the fourth order effective Hamiltonian in this subspace, see Appendix~\ref{app:effective}, can be written as
	\begin{align}
	\label{eq:fourth_order}
	H_{\rm eff}^{(4)}=-2\alpha^4\hbar\delta (2a^\dagger a+1) \left[\sigma_+^{(1)}\sigma_-^{(2)}+
	\sigma_-^{(1)}\sigma_+^{(2)}\right].
	\end{align}
	The ratio between this fourth-order contribution and the second-order term~\eqref{eq:second_order} in the protected subspace is given by $2\alpha^2(2a^\dagger a+1)$. Therefore, the fourth order is usually neglected in the previous literature for the case of ground-state cooling of the oscillator, i.e., low average phonon occupations, but this approximation may be inaccurate for cases involving thermal phonons~\cite{Sor99,Klaus99}. 
	
	\subsection{Dynamics in the protected subspace}
	\label{sec:fidelity1}
	Combining Eqs.~\eqref{eq:second_order} and~\eqref{eq:fourth_order}, we can rewrite the effective Hamiltonian in the protected subspace up to fourth order in $\alpha$ in the simple form
	\begin{align}
	\label{eq:effective_protected}
	H_{\rm eff}=\hbar\delta f(a^\dagger a)\varsigma_x
	\end{align}
	with the operator $\varsigma_x=|1\rangle\langle 0|+|0\rangle\langle 1|$ and the function
	\begin{align}
	f(x)=\alpha^2-2\alpha^4(2x+1).
	\end{align}
	For the case of an oscillator with a fixed initial phonon number, i.e., an initial oscillator Fock state $|n\rangle$, this Hamiltonian directly implies Rabi oscillations between $|0\rangle$ and $|1\rangle$ with a modified Rabi frequency $2f(n)\delta$. However, a more realistic initial condition is when the oscillator is in thermal equilibrium at temperature $T$. The density operator of the oscillator can then be written as
	\begin{align}
	\mu_{\rm th}=(1-q)q^{a^\dagger a},
	\end{align}
	where the Boltzmann factor $q = \exp\bla{-\hbar \nu/k_{\rm B} T}$ is related to the mean thermal occupation of the mechanical oscillator, i.e. $\bar{n}=[\exp(\hbar\nu/k_{\rm B}T)-1]^{-1}$, via $q=\bar{n}/(\bar{n}+1)$. If the spins are initially prepared in the state $|0\rangle$, the effective time evolution of the full system's density operator $\varrho$ is formally given by
	\begin{align}
	\varrho(t)=e^{-iH_{\rm eff}t/\hbar}\mu_{\rm th}|0\rangle\langle 0|e^{iH_{\rm eff}t/\hbar}.
	\end{align}
	Using 
	$\exp[i\lambda\varsigma_x]|0\rangle=\cos(\lambda)|0\rangle+i\sin(\lambda)|1\rangle$ and the fact that $f(a^\dagger a)$ naturally commutes with $\mu_{\rm th}$ yields the state
	\begin{align}
	\varrho(t)=\mu_{\rm th}\Big[&\cos^2(f(a^\dagger a)\delta t)|0\rangle\langle 0|+\sin^2(f(a^\dagger a)\delta t)|1\rangle\langle 1|\nonumber\\
	&-\frac{1}{2}\sin(2f(a^\dagger a)\delta t)\varsigma_y\Big]
	\end{align}
	with the operator $\varsigma_y=i|1\rangle\langle 0|-i|0\rangle\langle 1|$. Performing the partial trace over the oscillator degrees of freedom, denoted by ${\rm Tr}_{\rm M}\{\cdot\}$, leads to a reduced spin density operator $\rho(t)={\rm Tr}_{\rm M}\{\varrho(t)\}$. Employing the geometric series to obtain 
	\begin{align}
	\label{eq:exponential_sum}
	\sum_{n=0}^\infty x^n e^{iny}=\frac{1-xe^{-iy}}{1-2x\cos y+x^2}
	\end{align}
	and taking the real and imaginary part of this relation yields the result
	\begin{align}
	\label{eq:reduced_dynamics}
	\rho(t)=F(t)|0\rangle\langle 0|+[1-F(t)]|1\rangle\langle 1|+S(t)\varsigma_y.
	\end{align}
	\begin{figure}[t!]
		\begin{center}
			\includegraphics[width=0.95\linewidth]{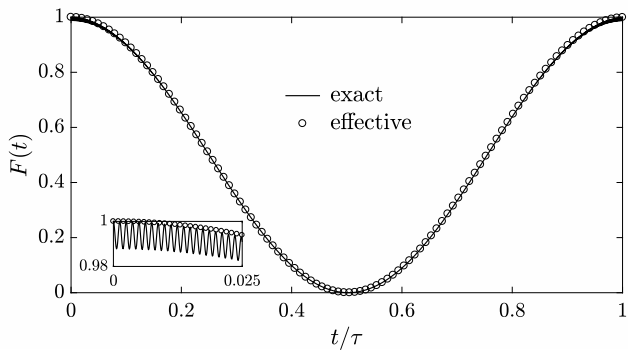}
		\end{center}
		\caption{Comparison between the fidelity $F(t)$ of the initial spin state $|0\rangle$ under the exact Hamiltonian~\eqref{eq:Tavis-Cummings} and the effective Hamiltonian~\eqref{eq:effective_protected}. The mean thermal occupation of the oscillator is $\bar{n}=2$ and the coupling to detuning ratio is $\alpha=1/40$. The time is given in units of the Rabi period $\tau=2\pi/\bar{g}$.}
		\label{Fig.lab.2}
	\end{figure}
	Here, we have introduced the fidelity $F(t)=1/2+C(t)$ to find the initial two-spin state $|0\rangle$ and the coherence $S(t)$. The functions $C(t)$ and $S(t)$ are, respectively, given by
	\begin{gather}
	\label{eq:cosine}
	C(t)=\frac{1-q}{2}\frac{\cos\big(\bar{g}t\big)-q\cos\big(2[\alpha^2\!+\!2\alpha^4]\delta t\big)}{1-2q\cos(8\alpha^4\delta t)+q^2},\\
	\label{eq:sine}
	S(t)=-\frac{1-q}{2}\frac{\sin\big(\bar{g}t\big)-q\sin\big(2[\alpha^2+2\alpha^4]\delta t\big)}{1-2q\cos(8\alpha^4\delta t)+q^2}
	\end{gather}
	with the effective Rabi frequency $\bar{g}=2(\alpha^2-2\alpha^4)\delta$. 
	
	In Fig.~\ref{Fig.lab.2}, we show the time evolution of the spin state $\ket{0}$ that lies in the protected subspace, and thus suffers only from fourth-order thermal effect of the mechanical oscillator, which is here assumed to be in a thermal state $\mu_{\rm th}$ with a mean phonon number $\bar{n}=2$. For the coupling ratio we chose $\alpha=1/40$. The thermal effect of the mechanical oscillator on the spins is significantly reduced, and the requirement on the cooling of the mechanical oscillator to obtain high fidelities can thus be relaxed. 
	
	The deviation of the approximate time evolution from the exact one, e.g., the absence of the small oscillations around the Rabi oscillations in the approximate time evolution, as shown in the inset of Fig.~\ref{Fig.lab.2}, has two origins: firstly, the truncation of the dispersive transform of the Hamiltonian, which leads to negligible discrepancies of the order $(2\bar{n}+1)^2\alpha^6$; secondly, the fact that the initial state has not been transformed to the dispersive frame, see Appendix~\ref{app:effective}.
	
	\section{Entangling gate via a thermal mechanical oscillator}
	In the above section, it has been shown that the dynamics in the protected subspace, spanned by $\{ \ket{0},\ket{1}\}$, is highly resistant against the thermal effects of the mechanical oscillator. In order to construct a robust entangling gate $U$, we therefore choose to encode the logical qubit in the two states $\ket{0}$ and $\ket{A}$, i.e., we use two spins to encode one logical qubit. Apart from $\ket{0}$ we use the state $|A\rangle=|a\rangle |a\rangle$, where the states $|a\rangle$ are decoupled from $|+\rangle$ and $|-\rangle$ [see Fig.~\ref{Fig.lab.1}(b)]. In this section, we show how to use the system introduced above to implement an entangling gate between two such logical qubits according to
	\begin{eqnarray}
	\label{eq:phase_flip}
	|0\rangle_{12}|0\rangle_{34} &\rightarrow& -|0\rangle _{12}|0\rangle_{34},\\
	|0\rangle_{12}|A\rangle_{34} &\rightarrow& |0\rangle _{12}|A\rangle_{34},\\
	|A\rangle_{12}|0\rangle_{34} &\rightarrow &|A\rangle_{12} |0\rangle_{34},\\
	|A\rangle_{12}|A\rangle_{34} &\rightarrow &|A\rangle_{12} |A\rangle_{34}.
	\end{eqnarray}
	Here, the notation $|\,\cdot\,\rangle_{jk}$ stands for the state of the logical qubit formed by the two physical qubits $j$ and $k$. The crucial minus sign in Eq.~\eqref{eq:phase_flip} arises due to the phase flip of the state of the state $|0\rangle_{23}$ after one Rabi period. It is therefore the fidelity of this Rabi oscillation, which is discussed in detail in Secs.~\ref{sec:fidelity1} and~\ref{sec:fidelity2}, that mainly determines the overall gate fidelity.
	
\begin{figure}[t!]
		\flushleft{\hspace{1ex} (a)}\vspace{-2.3ex}
		\begin{center}
			\hspace{3ex}\includegraphics[width=0.9\linewidth]{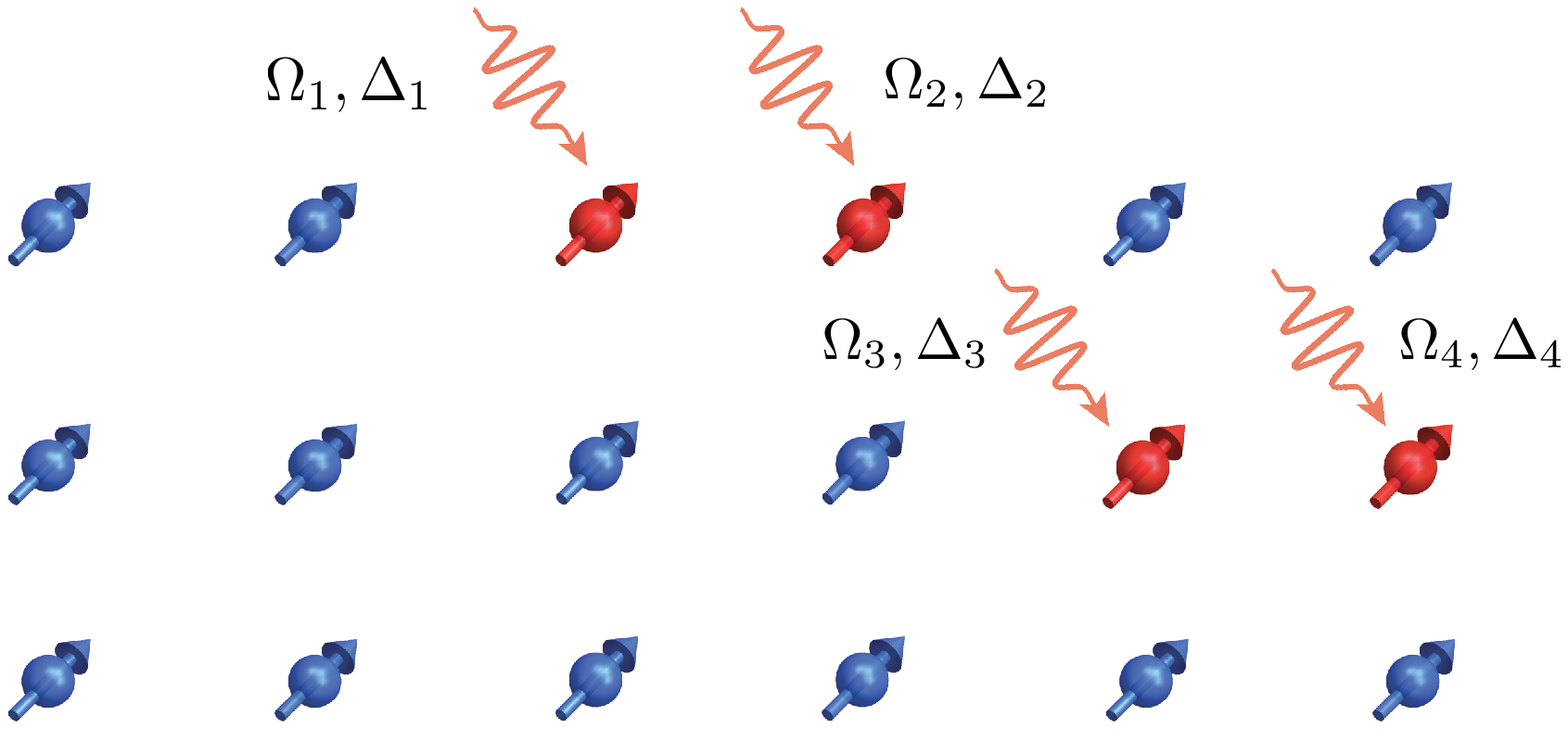}
		\end{center}\vspace{-0.0ex}
		\flushleft{\hspace{1ex} (b)}\vspace{-2.3ex}
		\begin{center}
			\includegraphics[width=0.95\linewidth]{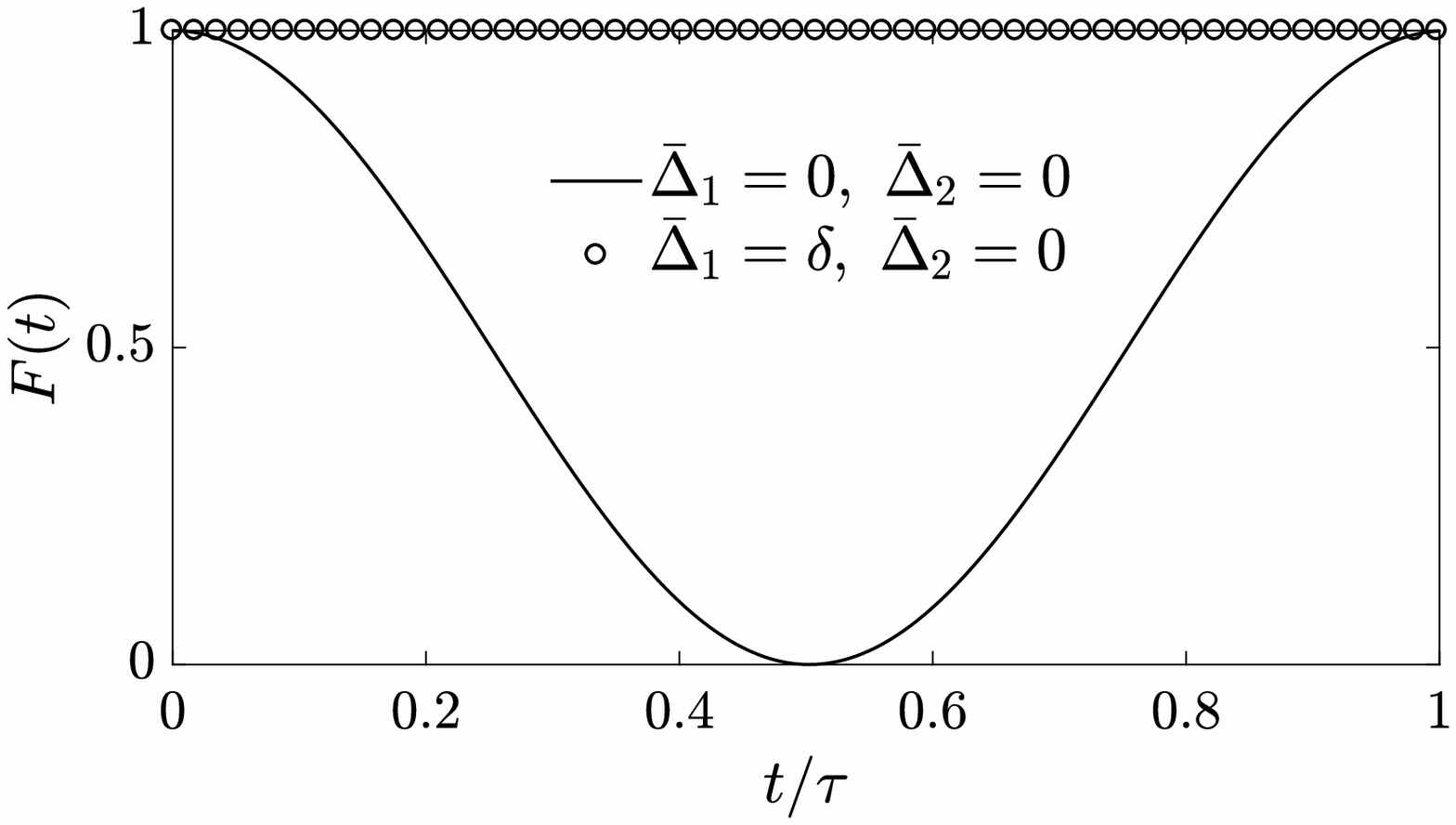}
		\end{center}
		\caption{(a) Architecture of a scalable spin array with one logical qubit encoded in two spins (marked in red). The coupling between two logical qubits is implemented using driving fields with suitable amplitudes $\Omega_k$ and detunings $\Delta_k$. Spins without drive are decoupled from each other.
			(b) Rabi oscillations between the two-qubit states $|0\rangle$ and $|1\rangle$ under resonance condition (solid line), i.e. $\bar{\Delta}_1=\bar{\Delta}_2$, and under large detuning (dots), i.e., $\bar{\Delta}_1 = \delta$ and $\bar{\Delta}_2 = 0$. The effective coupling ration is $\alpha=1/40$ and the thermal occupation of the oscillator is $\bar{n}=2$.}	
		\label{Fig.lab.3}
	\end{figure}

	We note that if for the ac Stark shift term in Eq.~\eqref{eq:effective} the condition $\bar{\Delta}_j=\bar{\Delta}_k$ is fulfilled, then the Rabi frequency between the states $|0\rangle_{jk}$ and $|1\rangle_{jk}$, up to fourth order in $\alpha$ and in the absence of thermal effects, is given by $\bar{g}$, see Eqs.~\eqref{eq:reduced_dynamics} and~\eqref{eq:cosine} for $q=0$. In contrast, the transition between $|0\rangle_{jk}$ and $|1\rangle_{jk}$ will be effectively suppressed if $\vert\bar{\Delta}_j-\bar{\Delta}_k \vert \gg \bar{g} $. Therefore, we choose the amplitudes and frequencies of the driving fields such that 
	\begin{gather}
	\bar{\Delta}_2=\bar{\Delta}_3=0
	\end{gather}
	and
	\begin{gather}
	\vert\bar{\Delta}_j-\bar{\Delta}_k \vert \gg \bar{g} 
	\end{gather}
	for all other combinations of $j$ and $k$, with $j,k=1,2,3,4$, which includes the conditions on the detunings of the first and the fourth spin. In this way, it can be seen that the engineered transition only happens for the spins 2 and 3, see Fig.~\ref{Fig.lab.3}. 
	If the two logical qubits are initially prepared in the state 
	\begin{align}
	|\psi(0)\rangle=\frac{1}{2}\Big[\big(|0\rangle_{12}+|A\rangle_{12}\big)\otimes \big(|0\rangle_{34}+|A\rangle_{34}\big)\Big]
	\end{align}
	after the time $\tau=2\pi/\bar{g}$, the state evolves into the maximally entangled state
	\begin{align}
	|\psi(\tau)\rangle=\frac{1}{2}\Big[|0\rangle_{12}&\otimes\big(|A\rangle_{34}-|0\rangle_{34}\big)\nonumber\\
	+|A\rangle_{12}&\otimes\big(|A\rangle_{34}+|0\rangle_{34}\big)\Big].
	\end{align}
	The initial state $|\psi(0)\rangle$ can be prepared from the state $|a\rangle_1|a\rangle_2$ using the following steps: (i) separately addressing the $|a\rangle$-$|g\rangle$- and $|a\rangle$-$|e\rangle$ transitions of the two spins to generate $|0\rangle_{12}=|-\rangle_1|+\rangle_2$. (ii) Since the effective Hamiltonian induces oscillations between $|0\rangle_{12}$ and $|1\rangle_{12}$, we can prepare the state $|0\rangle_{12}+|1\rangle_{12}=|e\rangle_1|e\rangle_2-|g\rangle_1|g\rangle_2$. (iii) Now driving the  $|g\rangle$-$|a\rangle$ of both spins creates $|e\rangle_1|e\rangle_2+|A\rangle_{12}$, from which one can prepare $|0\rangle_{12}+|A\rangle_{12}$. The same applies to the spins $3$ and $4$.
	
	In order to validate the robustness of such an entangling gate, as affected by the mean thermal phonon number $\bar{n}$, we can evaluate the fidelity $F(t)$, as given by Eq.~\eqref{eq:reduced_dynamics}, after the approximate Rabi period $\tau=2\pi/\bar{g}$. Expanding $C(\tau)$, defined in Eq.~\eqref{eq:cosine}, up to fourth order in $\alpha$ yields the gate fidelity
	\begin{eqnarray}
	F(\tau)=1-16\pi^2\bar{n}(2\bar{n}+1)\alpha^4+O(\alpha^6).
	\end{eqnarray}
	Therefore, the approximate infidelity of the entangling gate due to thermal effects of the mechanical oscillator reads
	\begin{equation}
	\label{eq:infidelity}
	\delta F_{\rm th} = 16\pi^2\bar{n}(2\bar{n}+1)\alpha^4.
	\end{equation}
	It can be seen that even for $\bar{n}=10$ and $\alpha= 1/40$ we have $\delta F_{\rm th}\approx 1.3\%$, i.e., for a thermal state occupation number $\bar{n} = 10$, it is still possible to reach a gate fidelity as high as $99\%$.
	For comparison, we also calculate the infidelity of an entangling gate without adopting our scheme, which is estimated as $\delta F_{\rm th}'=\bar{n}/(2 \bar{n}+1)$, see Appendix~\ref{app:gate_fidelity}. It thus requires the mechanical oscillator to be very close to the ground state in order to achieve a high gate fidelity. For example, it requires $\bar{n} < 0.125$ in order to achieve a gate fidelity above $90\%$, which implies a great challenge in experiments.
	
	\section{Analysis of decoherence effects}
	The main sources of decoherence in this hybrid system of solid-state spins and mechanical oscillator are the dephasing of the spins and the damping of the oscillator by its thermal environment. The dissipative dynamics under these decoherence effects can be described by the master equation
	\begin{equation}
	\begin{split}
	\frac{\partial}{\partial t}\varrho(t)=\frac{1}{i\hbar}[H,\varrho(t)]+ \mathcal{D}_{\rm S}\varrho(t) + \mathcal{D}_{\rm M}\varrho(t),
	\end{split}
	\end{equation}
	where $\mathcal{D}_{\rm S}$ represents the intrinsic pure dephasing of the spins at a rate $\Gamma/2$,  as given by
	\begin{eqnarray}
	\label{eq:dephaser}
	\mathcal{D}_{\rm S}\varrho =\frac{\Gamma}{2}\left(\sigma_x^{(1)}\varrho\sigma_x^{(1)}
	+\sigma_x^{(2)}\varrho\sigma_x^{(2)}
	- 2\rho\right).
	\end{eqnarray}
	In the case of NV centers, this dephasing has its origin in magnetic noise from nearby nuclear spins and coupling to lattice phonons. We remark that we have previously exchanged the $\sigma_z$ with the $\sigma_x$ basis and that thereby the pure dephasing in the original $\sigma_z$-basis also comprises a population decay in the $\sigma_x$ basis. The last term including $\mathcal{D}_{\rm M}$ describes the relaxation of the mechanical oscillator at a rate $\gamma$ given by
	\begin{align}
	\mathcal{D}_{\rm M}\varrho=&\frac{\gamma}{2}(\bar{n}+1)\left(2a\varrho a^\dagger-a^\dagger a\varrho-\varrho a^\dagger a\right)\nonumber\\
	&+\frac{\gamma}{2}\bar{n}\left(2a^\dagger\varrho a-aa^\dagger\varrho-\varrho aa^\dagger\right)
	\end{align} 
	We further note that in this description we neglected contributions to the decoherence and decay that arise during the transformation to the dispersive frame~\cite{Boissonneault09,Bennett13,Govia15}. Discarding these additional contributions is valid for $\Gamma,\gamma(\bar{n}+1)\ll \delta$. Including the thermal dissipation of the oscillator effectively leads to an additional collective relaxation of the spins at a rate $\alpha^2\gamma$. This relaxation and dephasing induced by the mechanical oscillator adds an additional source of spin dephasing~\cite{Clerk07,Rigetti12}, which may be neglected due to the smallness of $\alpha$. 
	
	The driving field plays a role of continuous dynamical decoupling. Thus, the effective decoherence rate $\Gamma$ can generally be suppressed by  two or three orders of magnitude as compared with the pure dephasing rate \cite{Rabl09,Farf17,Cai_NJP_2012,Cohen16}. Given a dephasing rate of 100 kHz, e.g. for NV center spins in diamond, the effective decoherence rate can be reduced to $\sim 1$ kHz, which may be much smaller than the coupling strength $g$.

	\subsection{Intrinsic spin dephasing}	
\begin{figure}[t!]
		\flushleft{\hspace{1ex} (a)}\vspace{-2.3ex}
		\begin{center}
			\includegraphics[width=0.95\linewidth]{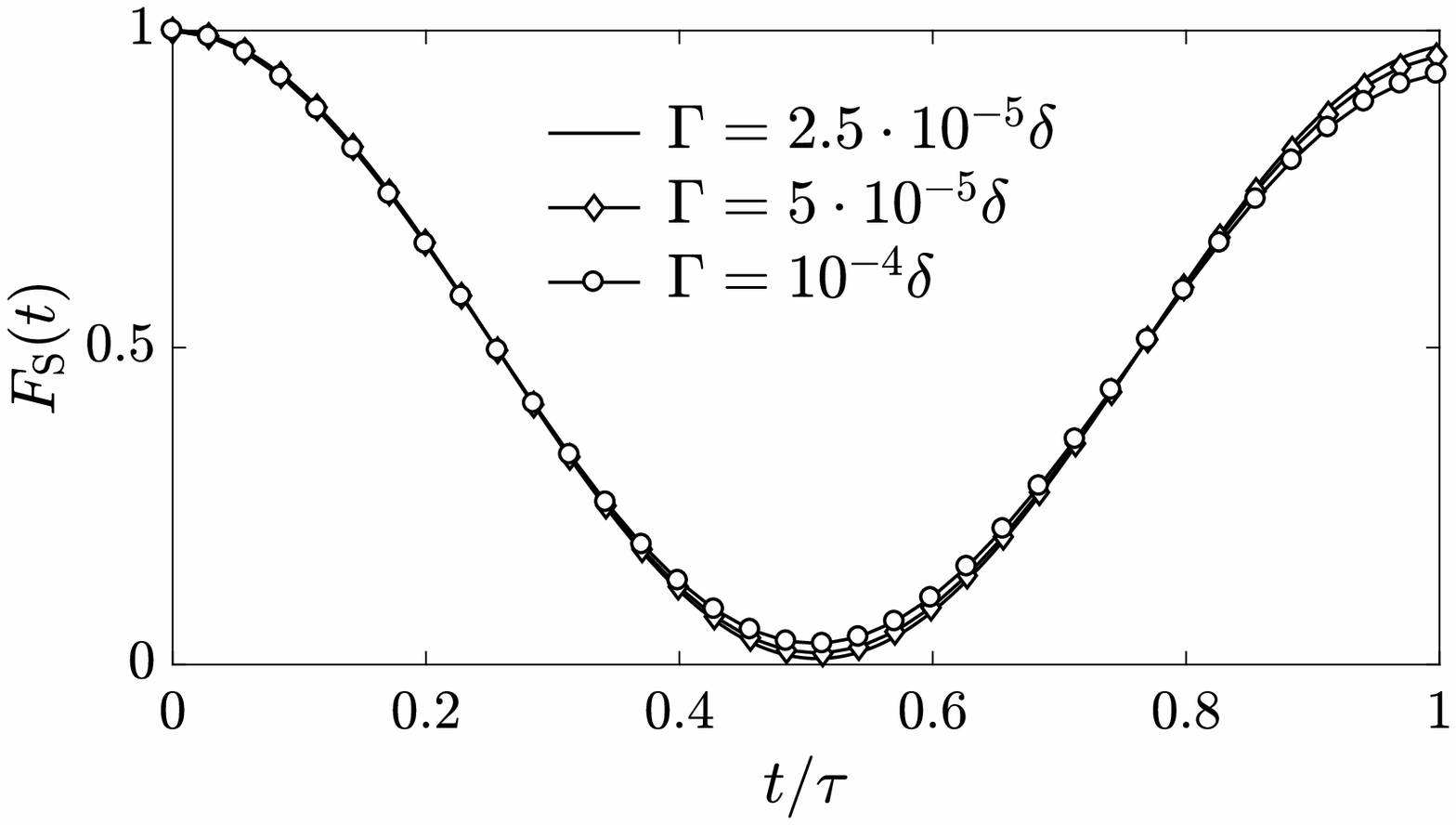}
		\end{center}\vspace{-0.0ex}
		\flushleft{\hspace{1ex} (b)}\vspace{-2.3ex}
		\begin{center}
			\includegraphics[width=0.95\linewidth]{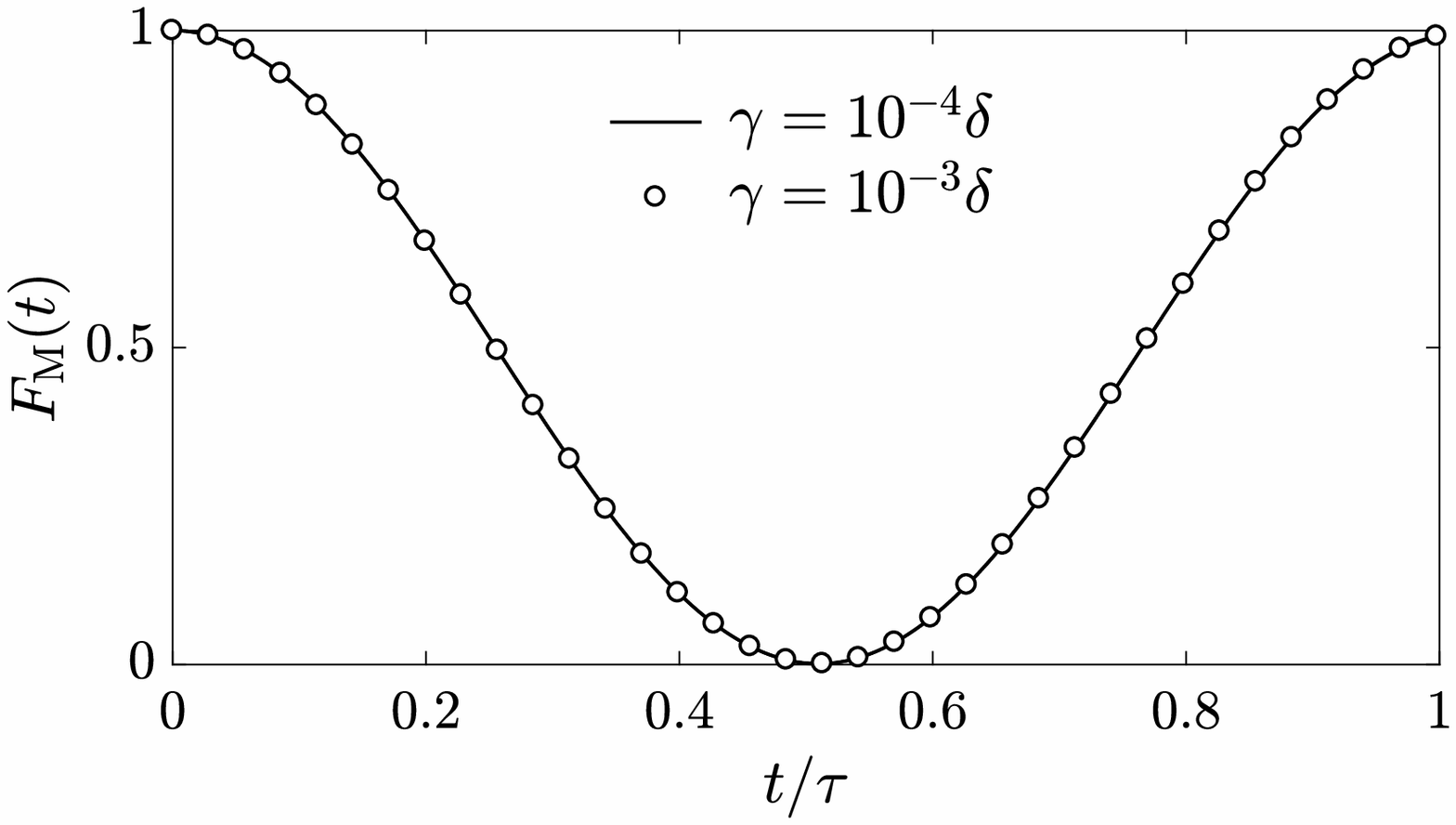}
		\end{center}
		\caption{\label{fig:spin_dephasing}(a) Fidelity of the initial spin state $|0\rangle$ under spin dephasing for the dephasing rates $\Gamma=2.5\cdot10^{-5}\delta$ (solid), $\Gamma=5\cdot10^{-5}\delta$ (diamonds), and $\Gamma=10^{-4}\delta$ (circles). (b) Fidelity of the initial spin state $|0\rangle$ under mechanical damping with the two rates $\gamma=10^{-4}\delta$ (solid) and $\gamma=10^{-3}\delta$ (circles). For both (a) and (b), the mean thermal occupation of the oscillator is $\bar{n}=2$ and we chose the coupling ratio $\alpha=1/20$. The time is given in units of the Rabi period $\tau$.}	
	\end{figure}

	In order to highlight the individual effects of the two decoherence channels on the gate fidelity, we will treat them separately here, starting with the case of only spin dephasing, i.e. $\gamma=0$. Although it is possible, as we will see, to derive an analytic expression for the time evolution of the initial state $\varrho(0)=\mu_{\rm th}|0\rangle\langle 0|$, the dephasing given by the superoperator~\eqref{eq:dephaser} lifts the confinement of the dynamics to the protected subspace spanned by $|0\rangle$ and $|1\rangle$. The dephasing additionally populates the density operator elements $|G\rangle\langle G|$ and $|E\rangle\langle E|$, with $|G\rangle=|-\rangle_1|-\rangle_2$ and $|E\rangle=|+\rangle_1|+\rangle_2$. In Appendix~\ref{app:spin_dephasing} we use the spectral decomposition of the Liouville operator~\cite{Briegel93,Barnett00} to show that the reduced spin density operator has the form
	\begin{align}
	\label{eq:spin_dephasing_density}
	\rho(t)=&F_{\rm S}(t)|0\rangle\langle 0|+\Big[1-F_{\rm S}(t)-\frac{1}{4}\big(1-e^{-2\Gamma t}\big)\Big]|1\rangle\langle 1|\nonumber\\
	&+e^{-\Gamma t}S(t)\varsigma_y+\frac{1}{4}\big(1-e^{-2\Gamma t}\big)\big[|E\rangle\langle E|+|G\rangle\langle G|\big],
	\end{align}
	where $F_{\rm S}(t)$ is the fidelity under spin dephasing given by
	\begin{align}
	\label{eq:fidelity_total}
	F_{\rm S}(t)=\frac{1}{4}\big(1+e^{-2\Gamma t}\big)+e^{-\Gamma t}C(t)
	\end{align}
	with the same functions $C(t)$ and $S(t)$ as defined in Eqs.~\eqref{eq:cosine} and~\eqref{eq:sine}. The steady state of the dynamics under spin dephasing is the fully mixed two-qubit state, yielding the long-time limit $F_{\rm S}(t)\to 1/4$. Figure~\ref{fig:spin_dephasing}(a) shows the impact of the spin dephasing on the fidelity for three different values of $\Gamma$ for the parameters $\bar{n}=2$ and $\alpha=1/20$. 
	
	\subsection{Mechanical damping}	
	Also for the case of strictly mechanical damping, i.e. $\Gamma=0$, it is possible to derive the dynamics of the initial state $\varrho(0)=\mu_{\rm th}|0\rangle\langle 0|$ analytically with the same methods, as shown in App.~\ref{app:mechanical_damping}. The time evolution of the reduced spin density matrix $\rho(t)$ reads
	\begin{align}
	\label{eq:mechanical_damping_density}
	\rho(t)=&F_{\rm M}(t)|0\rangle\langle 0|+[1-F_{\rm M}(t)]|1\rangle\langle 1|+S_{\rm M}(t)\varsigma_y,
	\end{align}
	where the fidelity $F_{\rm M}(t)$ and the coherence $S_{\rm M}(t)$ under mechanical damping are given by
	\begin{gather}
	\label{eq:fidelity_mec_damp}
	F_{\rm M}(t)=\frac{1}{2}+\frac{1}{2}{\rm Re}\{Y(t)\},\\
	\label{eq:coherence_mec_damp}
	S_{\rm M}(t)=-\frac{1}{2}{\rm Im}\{Y(t)\}
	\end{gather}
	with the function $Y(t)$ defined in the appendix, see Eq.~\eqref{eq:A}. In Fig.~\ref{fig:spin_dephasing}(b) we show the fidelity under mechanical damping for the same parameters, $\bar{n}=2$ and $\alpha=1/20$, for two values of the damping rate $\gamma$. Here, it is apparent that the direct influence of the mechanical relaxation on the entangling gate is rather small and can thereby be neglected for the evaluation of the gate fidelity, a fact also found in a different proposal for an entangling gate using a thermal oscillator~\cite{Schuetz17}.
	
	\subsection{Fidelity estimation}
	\label{sec:fidelity2}
	As it can be seen in Fig.~\ref{fig:spin_dephasing}(b), the fidelity under mechanical damping is almost unity due to the smallness of the parameter $\alpha$, since the effective spin relaxation rate induced by the dissipation of the mechanical oscillator is given by $\Gamma_{\rm s} = \alpha^2\gamma $. We can thus neglect this contribution and estimate the infidelity induced by thermal effects and spin dephasing from Eqs.~\eqref{eq:fidelity_total}.
	The infidelity
	\begin{equation}
	\begin{split}
	\delta F\approx 1 - F_{\rm S}(\tau)
	\end{split}
	\end{equation}
	of the above entangling gate can then be estimated as 
	\begin{equation}
	\delta F\approx 16\alpha^4\pi^2\bar{n}(2\bar{n}+1)+ \frac{\pi\Gamma}{(\alpha^2-2\alpha^4)\delta},
	\end{equation}
	where the first term comes from the thermal effect and the second term arises from the spin dephasing.
	
	It can be seen that the thermal effect of the mechanical oscillator, as shown in the first term in the above equation, is suppressed by a factor of $O(\alpha^4)$. The spin dephasing is suppressed by continuous dynamical decoupling, and thus its influence on the fidelity of the entangling gate can also be efficiently suppressed. In Fig.~\ref{fig:spin_dephasing}(a), we verify our estimation of the gate infidelity of the Rabi oscillation between the two-qubit states $|0\rangle$ and $|1\rangle$ assuming decoherence parameters $\Gamma$ and $\gamma$ that are realistic in experiments. It can also be seen that the gate fidelity can reach up to $90\%$, achieved even with a relative high phonon number and under the influence of the dissipation of the mechanical oscillator and the spin dephasing.
	
	\section{Discussion and conclusion}
	\label{sec:conclusion}
	The present idea may be applied to general hybrid systems of (artificial) atoms and a mechanical oscillator.  As a specific example, we assume a mechanical oscillator with a resonance frequency $\nu = (2\pi) 1$ MHz~\cite{Muschik14}, and a coupling strength $g=(2\pi) 100$ kHz. These parameters are achievable for a hybrid system of NV center spins in diamond and a mechanical oscillator, such as cantilevers or nanotubes~\cite{Rabl09,Peng16}. We choose the value of the detuning as $\delta = (2\pi) 4$ MHz to satisfy the condition $\alpha=1/40\ll 1$. For a mechanical quality factor of $Q=4\cdot10^3$, which is even lower than in experimental realizations~\cite{Weber14,Will17,Christoph16}, we have the dissipation rate $\gamma = (2\pi)10 $ kHz. For NV center spins, the extended coherence time can reach up to millisecond, and the effective decoherence rate can be estimated as $\Gamma \sim (2\pi)100$ Hz with continuous dynamical decoupling methods~\cite{Naydenov11}. Using these parameters, our results show that the entangling gate fidelity can reach $94\%$ even when the mechanical oscillator is in a thermal state with an average phonon number of $\bar{n} = 10$.
	
	To summarize, we have proposed a new method to drastically suppress the thermal effect of a mechanical oscillator in a hybrid system where the spin-spin interaction is mediated by a harmonic oscillator. With the mechanical oscillator as a mediator, it is possible to implement a robust entangling gate between two solid-state spin qubits. Our scheme combines the advantages of continuous dynamical decoupling and decoherence-free subspaces. With our detailed investigation we show that the fidelity of the entangling gate can be sufficiently high even with a relatively large average phonon number. The scheme thus provides a way towards the implementation of scalable quantum information processing in solid-state architectures.
	
	\section{Acknowledgments}
	We thank Haibin Liu and Itsik Cohen for helpful discussions and suggestions. This work
	is supported by the National Natural Science Foundation of China (Grant No. 11690030, 11690032, 11574103) and the National
	Young 1000 Talents Plan.
	
	\appendix
	\section{Effective Hamiltonian}
	\label{app:effective}
	We start with the Hamiltonian~\eqref{eq:Tavis-Cummings}, where we consider the resonant case $\Delta_k=0$ for reasons of convenience. Additionally, we consider a uniform coupling $g_k=g$ and detuning $\delta_k=\delta$. In order to cast the following expansion in a more comprehensive form we introduce the collective spin operators $S_z=\sum_k\sigma_{z}^{(k)}$ and $S_\pm=\sum_k\sigma_{\pm}^{(k)}$ as well as the two operators $J_\pm=aS_+ \pm a^\dagger S_-$. With the coupling ratio $\alpha=g/\delta$ defined in the main text, we can write Hamiltonian~\eqref{eq:Tavis-Cummings} in the frame rotating at the oscillator frequency $\nu$ as 
	\begin{align}
	H =\hbar\delta\left[\frac{1}{2}S_z+ \alpha J_+\right].
	\end{align}
	The transformation to the dispersive frame is a frequently used tool to treat this kind of Hamiltonian in the regime $\alpha\ll 1$~\cite{Bennett13,Blais04,Govia16,Zueco09,Boissonneault09}. Although this method is well established, we nevertheless repeat the basic steps here for the sake of completeness. It is achieved by the unitary transform $\exp[\alpha J_-]$. Using the Baker-Campbell-Hausdorff formula, the Hamiltonian transforms according to
	\begin{align}
	\label{eq:Baker_Hausdorff}
	e^{\alpha J_-}He^{-\alpha J_-}=\sum_{n=0}^\infty\frac{\alpha^n}{n!}[J_-,H]_n
	\end{align}
	with the nested commutators $[X,Y]_{n+1}=[X,[X,Y]_n]$, where $[X,Y]_0=Y$. This allows to conveniently truncate the transform at a given order of $\alpha$. Realizing that $[J_-,S_z/2]=-J_+$ leads to a vanishing first order and allows us to separate the remaining orders of $\alpha$ as
	\begin{align}
	e^{\alpha J_-}He^{-\alpha J_-}=\frac{\hbar\delta}{2}S_z+\sum_{n=2}^\infty H_{\rm eff}^{(n)}
	\end{align}
	with the $n$th order effective Hamiltonian
	\begin{align}
	\label{eq:eff}
	H_{\rm eff}^{(n)}=\alpha^n\hbar\delta\frac{n-1}{n!}[J_-,J_+]_{n-1}
	\end{align}
	for $n\geq 2$. The second order effective Hamiltonian simply reads
	\begin{align}
	H_{\rm eff}^{(2)}=\frac{\alpha^2\hbar\delta}{2}[(2a^\dagger a+1)S_z+S_+S_-+S_-S_+].
	\end{align}
	Evaluating the next commutators of Eq.~\eqref{eq:eff} then yields the fourth order effective Hamiltonian~\eqref{eq:effective_protected} in the protected subspace.
	
	The same effective Hamiltonian can also be derived using the Magnus expansion~\cite{Blanes09} or similar methods~\cite{James07}. As mentioned in the main text, for the case of a non-vanishing detuning $\Delta_k\gg \Omega_k$ one obtains an additional ac Stark shift due to the driving field, see Eq.~\eqref{eq:effective}. On a further note, one has to keep in mind that although the Hamiltonian was transformed into the dispersive frame the contributions of the transform to the density operator are usually neglected. These contributions are rather small but manifest themselves in slight discrepancies when comparing the exact time evolution with the one in the dispersive frame, such as the small oscillations in the exact fidelity curve as highlighted in the inset of Fig.~\ref{Fig.lab.2}.
	
	\section{Gate fidelity without thermal protection}
	\label{app:gate_fidelity}
	Without thermal protection, i.e., if the logical qubits are encoded in single physical qubits, the thermal effect takes place in the second order of $\alpha$ term
	\begin{align}
	H_{\rm th}
	=\frac{\alpha^2\hbar\delta}{2}(2a^\dagger a+1)S_{z}.
	\end{align}
	In order to estimate the entangling gate fidelity in this case, we use single-qubit states which do not lie within the decoherence-free subspace. For the case of an oscillator Fock state $|n\rangle$, this leads to a spin phase flip 
	\begin{equation}
	\begin{split}
	|+\rangle|-\rangle &\rightarrow -|+\rangle|-\rangle, \\
	|+\rangle|a\rangle &\rightarrow \beta_n |+\rangle|a\rangle,\\
	|a\rangle|+\rangle &\rightarrow \beta_n^\ast |a\rangle|+\rangle,\\
	|a\rangle|a\rangle &\rightarrow  |a\rangle|a\rangle.
	\label{Equa.21}
	\end{split}
	\end{equation}
	Here, the phase $\beta_n$ is the phase accumulated by the state $|+\rangle$, which after the approximate flip time $\tau\approx \pi/\alpha^2\delta$ is given by
	\begin{equation}
	\begin{split}
	\beta_n &= e^{-i(2n+1)\pi/2}.
	\end{split}
	\end{equation}
	For the initial state $|\psi_n(0)\rangle =|\phi(0)\rangle|n\rangle$, with $|\phi(0)\rangle=\left(|+\rangle +|a\rangle\right)\otimes \left(|-\rangle +|a\rangle\right)/2$, we easily find the spin state
	\begin{align}
	|\phi_n(\tau)\rangle = \frac{1}{2}\big[-|+\rangle|-\rangle + \beta_n|+\rangle|a\rangle+\beta_n^\ast|a\rangle|-\rangle + |a\rangle|a\rangle\big].
	\end{align}
	To evaluate the effects of \textit{n} on this entangling process, we compare the state $|\phi_n(\tau)\rangle $ with $|\phi_0(\tau)\rangle$. For a thermal oscillator state with mean phonon number $\bar{n}$, the fidelity losses can then be evaluated by
	\begin{equation}
	\begin{split}
	\delta F_{\rm th}^{(2)}&=1-(1-q)\sum_{n=0}^{\infty}q^n\left| \langle \phi_n(\tau)|\phi_0(\tau)\rangle \right|^2
	= \dfrac{\bar{n}}{(2 \bar{n}+1)},
	\label{Equa.23}
	\end{split}
	\end{equation}
	which is greatly influenced by the thermal occupation $\bar{n}$ of the mechanical oscillator as compared to our thermally protected result given by Eq.~\eqref{eq:infidelity}.
	
	\section{Dynamics under spin dephasing}
	\label{app:spin_dephasing}
	The time evolution of the system undergoing spin dephasing obeys the master equation $\partial\varrho/\partial t=\mathcal{L}\varrho$ with the Liouville operator
	\begin{align}
	\mathcal{L}\varrho=&-i\delta\big[f(a^\dagger a)\big(\sigma_+^{(1)}\sigma_-^{(2)}+\sigma_-^{(1)}\sigma_+^{(2)}\big),\varrho\big]\nonumber\\
	&+\frac{\Gamma}{2}\big[\sigma_x^{(1)}\varrho\sigma_x^{(1)}+\sigma_x^{(2)}\varrho\sigma_x^{(2)}-2\varrho\big].
	\end{align}
	\subsection{Damping basis}
	To solve this equation, we employ the spectral decomposition of the Liouville operator $\mathcal{L}$, details on the application of this technique to solve master equations of the Lindblad type can be found in Refs.~\cite{Briegel93,Barnett00,Betzholz14,Torres14}. The mechanical initial state $\mu_{\rm th}$ is diagonal in the Fock basis, making it unnecessary to find the full eigensystem of $\mathcal{L}$. We therefore focus on the subset of eigenelements that are also diagonal. We make the ansatz
	\begin{align}
	\hat{\varrho}_{n,j}=|n\rangle\langle n|\hat\rho_j,
	\end{align}
	which upon substitution into the eigenvalue equation $\mathcal{L}\hat{\varrho}_{n,j}=\lambda_{n,j}\hat{\varrho}_{n,j}$ leads to the condition
	\begin{align}
	\lambda_{n,j}\hat\rho_j=&-if(n)\delta [\sigma_+^{(1)}\sigma_-^{(2)}+\sigma_-^{(1)}\sigma_+^{(2)},\hat\rho_j]\nonumber\\
	&+\frac{\Gamma}{2}\big[\sigma_x^{(1)}\hat\rho_j\sigma_x^{(1)}+\sigma_x^{(2)}\hat\rho_j\sigma_x^{(2)}-2\hat\rho_j\big].
	\end{align}
	The relevant eigenvalues and the corresponding right eigenelements solving this eigenvalue problem are summarized in the table below.
	\begin{table}[h!]
		\centering
		\def\arraystretch{1.5}
		\begin{tabular}{|c|}\hline
			$\lambda_{n,j}$ \\
			\hline\hline
			$\lambda_{n,1}=0$  \\ \hline
			$\lambda_{n,z}=-2\Gamma$  \\ \hline
			$\lambda_{n,\pm}=\mp2if(n)\delta-\Gamma$  \\
			\\ \hline
		\end{tabular} \begin{tabular}{|c|}\hline
			$\hat\rho_\lambda$ \\
			\hline\hline
			$\hat{\rho}_1=1/4$ \\ \hline
			$\hat{\rho}_z=\sigma_z^{(1)}\sigma_z^{(2)}/4$ \\ \hline
			$\hat{\rho}_\pm=(\sigma_z^{(1)}-\sigma_z^{(2)})/16$ \\ 
			$\pm i(\sigma_x^{(1)}\sigma_y^{(2)}-\sigma_y^{(1)}\sigma_x^{(2)})/16$ \\ \hline
		\end{tabular}
	\end{table}
	
	\noindent Here, the $1$ in the eigenelement $\hat\rho_1$ stands for the unity operator on the two-qubit Hilbert space. The corresponding left eigenelements are given by $\check{\varrho}_{n,j}=|n\rangle\langle n|\check\rho_j$ with $\check{\rho}_1=4\hat{\rho}_1$, $\check{\rho}_z=4\hat{\rho}_z$, and $\check{\rho}_\pm=16\hat{\rho}_\mp$. They can either be constructed by using the orthogonality relation ${\rm Tr}\{\check{\rho}_j\hat{\rho}_k\}=\delta_{j,k}$ or by solving the eigenvalue equation of the adjoint superoperator $\mathcal{L}^\ddagger$~\cite{Barnett00}. 
	\subsection{Time evolution}
	The time evolution of our initial state $\varrho(0)=\mu_{\rm th}|0\rangle\langle 0|$ is then formally given by
	\begin{align}
	\varrho(t)=\sum_{n=0}^\infty\sum_{j=1,z,\pm}c_{n,j}e^{\lambda_{n,j}t}|n\rangle\langle n|\hat{\rho}_j
	\end{align}
	with the expansion coefficients $c_{n,j}={\rm Tr}\{\check{\varrho}_{n,j}|n\rangle\langle n|\varrho(0)\}$. Their explicit form is easily derived and reads $c_{n,1}=(1-q)q^n$, $c_{n,z}=-c_{n,1}$, and $c_{n,\pm}=2c_{n,1}$. This leads to the state
	\begin{align}
	\varrho(t)=&\frac{1}{4}\mu_{\rm th}\Big(1+e^{-2\Gamma t}\sigma_z^{(1)}\sigma_z^{(2)}\Big)
	+\frac{1-q}{4}e^{-\Gamma t}\nonumber\\
	&\times\sum_{n=0}^\infty|n\rangle\langle n|\bigg\{\cos(2f(n)\delta t)\big[\sigma_z^{(1)}-\sigma_z^{(2)}\big]\nonumber\\
	&-\sin(2f(n)\delta t)\big[\sigma_x^{(1)}\sigma_y^{(2)}-\sigma_y^{(1)}\sigma_x^{(2)}\big]\bigg\}
	\end{align}
	which, upon taking the partial trace over the oscillator and using the identity~\eqref{eq:exponential_sum}, yields the reduced spin density operator~\eqref{eq:spin_dephasing_density}, which is given in terms of the states $|0\rangle$, $|1\rangle$, $|G\rangle$ and $|E\rangle$ in the main text.
	
	\section{Dynamics under mechanical damping}
	\label{app:mechanical_damping}
	The Liouville operator of the master equation $\partial\varrho/\partial t=\mathcal{L}\varrho$ describing the dynamics under mechanical damping has the form
	\begin{align}
	\mathcal{L}\varrho=-i\delta\big[f(a^\dagger a)\varsigma_x,\varrho\big]+\mathcal{D}_{\rm M}\varrho.
	\end{align}
	\subsection{Damping basis}
	Here, again we do not present the full eigensystem of $\mathcal{L}$ but merely sketch the derivation of the eigenelements relevant for separable initial states whose mechanical density operator is diagonal in the Fock basis~\cite{Betzholz?}. 
	The relevant subset can be grouped into two sets. For the eigenelements of the first set we make the ansatz
	\begin{align}
	\hat{\varrho}_{\mu,\pm,n}&=\frac{1}{2}\left[1\pm\varsigma_x\right]\hat{\mu}_n,\\
	\check{\varrho}_{\mu,\pm,n}&=\frac{1}{2}\left[1\mp\varsigma_x\right]\check{\mu}_n.
	\end{align}
	For the diagonal mechanical operators $\hat{\mu}_n$ we thereby find the eigenvalue problem $\mathcal{D}_{\rm M}\hat{\mu}_n=\Lambda_{\mu,\pm,n}\hat{\mu}_n$. This is nothing but the eigenvalue equation of the damped harmonic oscillator, whose solutions $\hat{\mu}_n$, alongside their left eigenelements $\check{\mu}_n$, with the corresponding eigenvalues $\Lambda_{\mu,\pm,n}=-n\gamma$, can be found in Ref.~\cite{Briegel93}.
	
	For the second set we make the ansatz
	\begin{align}
	\hat{\varrho}_{\eta,\pm,n}=&\frac{1}{4}\left[\varsigma_z\pm i\varsigma_y\right]\hat{\eta}_{\pm,n},\\
	\check{\varrho}_{\eta,\pm,n}=&\left[\varsigma_z\mp i\varsigma_y\right]\check{\eta}_{\pm,n}.
	\end{align}
	Substituting this ansatz into the eigenvalue equation of $\mathcal{L}$ for the eigenvalue $\Lambda_{\eta,\pm,n}$ yields the new eigenvalue problem 
	\begin{align}
	\mathcal{K}\hat\eta_{\pm,n}=[\Lambda_{\eta,\pm,n}\mp i\bar g+\gamma\bar{n}]\hat{\eta}_{\pm,n},
	\end{align}
	for the mechanical operators $\hat{\eta}_{\pm,n}$, where the superoperator $\mathcal{K}$ is defined by its action
	\begin{align}
	\mathcal{K}\eta=&\left[\mp 4i\alpha^4\delta-\frac{\gamma}{2}(2\bar{n}+1)\right]\big\{a^\dagger a,\eta\big\}\nonumber\\
	&+\gamma(\bar{n}+1)a\eta a^\dagger+\gamma\bar{n} a^\dagger\eta a,
	\end{align}
	with the anticommutator $\{\cdot,\cdot\}$. The eigenvalue equation of $\mathcal{K}$ can be solved by transforming it into the associated Fokker-Planck equation~\cite{risken}. Again, the left eigenelements can be derived by solving the same problem for the adjoint operator $\mathcal{K}^\ddagger$. As mentioned before, a detailed derivation will be given elsewhere, and we merely give the resulting eigenvalues and eigenelements
	\begin{align}
	\Lambda_{\eta,\pm,n}=&\pm i\bar{g}-n\gamma(A+2\xi)-\gamma(\xi-1),\\
	\hat{\eta}_{\pm,n}=&\frac{(A+2\xi)(\bar{n}+1-\xi)}{\bar{n}(\bar{n}+1)}\left[\frac{\bar{n}+1-\xi}{\bar{n}+1+A+\xi}\right]^n\times\nonumber\\
	&:L_n\left(\frac{A+2\xi}{\bar{n}+1}a^\dagger a\right)e^{-\frac{\xi}{\bar{n}+1}a^\dagger a}:,\\
	\check{\eta}_{\pm,n}=&:L_n\left(\frac{A+2\xi}{\bar{n}}a^\dagger a\right)e^{-\frac{\xi-1}{\bar{n}}a^\dagger a}:.
	\end{align}
	Here, $:\cdot:$ denotes normal ordering of the expression enclosed by colons, $L_n(x)$ are the Laguerre polynomials, and we introduced the abbreviations $\xi=\left[\sqrt{A^2+4B}-A\right]/2$, $A=[\pm 8i\alpha^4\delta-\gamma]/\gamma$, as well as $B=[\pm 8i\alpha^4(\bar{n}+1)\delta]/\gamma$.
	\subsection{Time evolution}
	With the necessary eigenelements at hand we can proceed to evaluate the time evolution of the system, formally given by
	\begin{align}
	\varrho(t)=\sum_{j=\mu,\eta}\sum_{k=\pm}\sum_{n=0}^\infty c_{j,k,n}e^{\Lambda_{j,k,n} t}\hat{\varrho}_{j,k,n}.
	\end{align}
	Evaluating the overlap of the initial state with eigenelements, viz. $c_{j,\pm,n}={\rm Tr}\{\check{\varrho}_{j,\pm,n}\varrho(0)\}$ for $j=\mu,\eta$, yields the state
	\begin{align}
	\varrho(t)=\frac{1}{2}\mu_{\rm th}+\left\{\sum_{n=0}^\infty\left[-\frac{A+\xi}{\xi}\right]^n\frac{e^{\Lambda_{\eta,+,n}t}}{\xi}\hat{\varrho}_{\eta,+,n}+{\rm H.c.}\right\}
	\end{align} 
	which upon taking the partial trace over the oscillator and evaluating the sum over $n$ gives the reduced spin density operator
	\begin{align}
	\rho(t)=\frac{1}{2}+\frac{1}{2}{\rm Re}\{Y(t)\}\varsigma_z-\frac{1}{2}{\rm Im}\{Y(t)\}\varsigma_y.
	\end{align}
	Here, we introduced the function $Y(t)$ we used in Eqs.~\eqref{eq:fidelity_mec_damp} an~\eqref{eq:coherence_mec_damp} of the main text, which is given by
	\begin{align}
	\label{eq:A}
	Y(t)=\frac{e^{(2i\alpha^2\delta+\gamma/2)t}}{e^{-(A+2\xi)\gamma t/2}+y\sinh\left([A+2\xi]\gamma t/2\right)}
	\end{align}
	with $y=2(\bar{n}+1+A+\xi)\xi^2/(\bar{n}+1)(A+2\xi)$.

		\end{document}